\documentclass[8.5pt,twoside,twocolumn]{article}
\usepackage[super,sort&compress,comma]{natbib} 
\usepackage[version=3]{mhchem}
\usepackage{balance}
\usepackage{times,mathptm}
\usepackage{sectsty}
\usepackage{graphicx} 
\usepackage{lastpage}
\usepackage[format=plain,justification=raggedright,singlelinecheck=false,font=small,labelfont=bf,labelsep=space]{caption} 
\usepackage{fancyhdr}
\begin{document} 

\thispagestyle{plain}
\fancypagestyle{plain}{
\fancyhead[L]{\includegraphics[height=8pt]{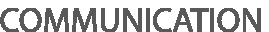}}
\fancyhead[C]{\hspace{-1cm}\includegraphics[height=20pt]{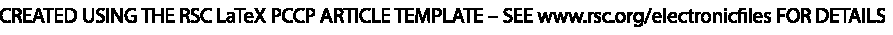}}
\fancyhead[R]{\hspace{10cm}\vspace{-0.25cm}\includegraphics[height=10pt]{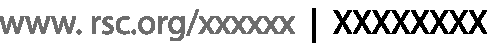}}
\renewcommand{\headrulewidth}{1pt}}
\renewcommand{\thefootnote}{\fnsymbol{footnote}}
\renewcommand\footnoterule{\vspace*{1pt}%
\hrule width 3.4in height 0.4pt \vspace*{5pt}} 
\setcounter{secnumdepth}{5}

\makeatletter 
\renewcommand\@biblabel[1]{#1}            
\renewcommand\@makefntext[1]%
{\noindent\makebox[0pt][r]{\@thefnmark\,}#1}
\makeatother 
\renewcommand{\figurename}{\small{Fig.}~}
\sectionfont{\large}
\subsectionfont{\normalsize} 

\fancyfoot{}
\fancyfoot[LO,RE]{\vspace{-7pt}\includegraphics[height=9pt]{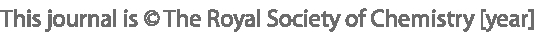}}
\fancyfoot[CO]{\vspace{-7.2pt}\hspace{12.2cm}\includegraphics{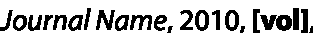}}
\fancyfoot[CE]{\vspace{-7.5pt}\hspace{-13.5cm}\includegraphics{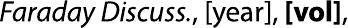}}
\fancyfoot[RO]{\footnotesize{\sffamily{1--\pageref{LastPage} ~\textbar  \hspace{2pt}\thepage}}}
\fancyfoot[LE]{\footnotesize{\sffamily{\thepage~\textbar\hspace{3.45cm} 1--\pageref{LastPage}}}}
\fancyhead{}
\renewcommand{\headrulewidth}{1pt} 
\renewcommand{\footrulewidth}{1pt}
\setlength{\arrayrulewidth}{1pt}
\setlength{\columnsep}{6.5mm}
\setlength\bibsep{1pt}

\twocolumn[
  \begin{@twocolumnfalse}
\noindent\LARGE{\textbf{Expansion and rupture of pH-responsive microcapsules$^\dag$}}
\vspace{0.6cm}

\noindent\large{\textbf{Sujit S. Datta,\textit{$^{a\ddag}$} Alireza Abbaspourrad,\textit{$^{a\ddag}$} and
David A. Weitz$^{\ast}$\textit{$^{a}$}}}\vspace{0.5cm}

\noindent\textit{\small{\textbf{Received Xth XXXXXXXXXX 20XX, Accepted Xth XXXXXXXXX 20XX\newline
First published on the web Xth XXXXXXXXXX 200X}}}

\noindent \textbf{\small{DOI: 10.1039/b000000x}}
 \end{@twocolumnfalse} \vspace{0.6cm}

  ]

\noindent\textbf{We study the deformations of pH-responsive spherical microcapsules -- micrometer-scale liquid drops surrounded by thin, solid shells -- under the influence of electrostatic forces. When exposed to a large concentration of NaOH, the microcapsules become highly charged, and expand isotropically. We find that the extent of this expansion can be understood by coupling electrostatics with shell theory; moreover, the expansion dynamics is well described by Darcy's law for fluid flow through the microcapsule shell. Unexpectedly, however, below a threshold NaOH concentration, the microcapsules begin to disintegrate, and eventually rupture; they then expand non-uniformly, ultimately forming large, jellyfish-like structures. Our results highlight the fascinating range of behaviors exhibited by pH-responsive microcapsules, driven by the interplay between electrostatic and mechanical forces. }
\section*{}
\vspace{-1cm}
\footnotetext{\dag~Electronic Supplementary Information (ESI) available: Experimental details, movies showing microcapsule expansion and rupture. See DOI: 10.1039/b000000x/}

\footnotetext{\textit{$^{a}$~Department of Physics and SEAS, Harvard University, Cambridge MA 02138, USA. Email: weitz@seas.harvard.edu}}
\footnotetext{\ddag~These authors contributed equally to this work.}

Microcapsules -- micrometer-scale liquid drops, each surrounded by a thin, solid shell -- are promising candidates for the encapsulation and controlled release of many technologically important active materials. These applications often require microcapsules to have unique mechanical properties, such as the ability to withstand large deformations.\cite{review} One common way of characterizing this behavior is to monitor how a microcapsule responds to forces exerted on its shell. These can be mechanical forces, externally imposed by poking, squeezing, or uniformly pressurizing the microcapsule.\cite{vernita,hans,fery,dattabuckling} Alternatively, these forces can be generated by physico-chemical modifications to the shell itself, such as charging it; under certain conditions, the repulsion between the charges on the shell can cause it to deform. Electrostatic forces arise in many applications of soft matter, and consequently, this approach is frequently used to induce deformations in a variety of bulk materials.\cite{estat1,estat2} Nevertheless, systematic investigations of how electrostatic forces deform microcapsules are scarce.\cite{swelling1,swelling2,swelling3} Thus, despite their prevalence in many real-world situations, a full understanding of how these forces affect microcapsules is lacking. 

In this Communication, we study the deformations of spherical microcapsules exposed to a pH stimulus. We choose NaOH as the stimulus; when exposed to this base, the microcapsule shells become highly charged. For large NaOH concentrations, the microcapsules expand isotropically. We find that the extent of the microcapsule expansion can be understood by coupling electrostatics with shell theory; moreover, the dynamics of this expansion is well described by Darcy's law for flow through the porous microcapsule shell. Surprisingly, below a threshold NaOH concentration, the microcapsules begin to disintegrate, and eventually rupture; they then expand non-uniformly, ultimately forming large, jellyfish-like structures. Our results thus highlight the rich behavior exhibited by microcapsules, driven by the interplay between electrostatic and mechanical forces. 

 \begin{figure*}
\begin{center}
\includegraphics[width=2.9in,angle=270]{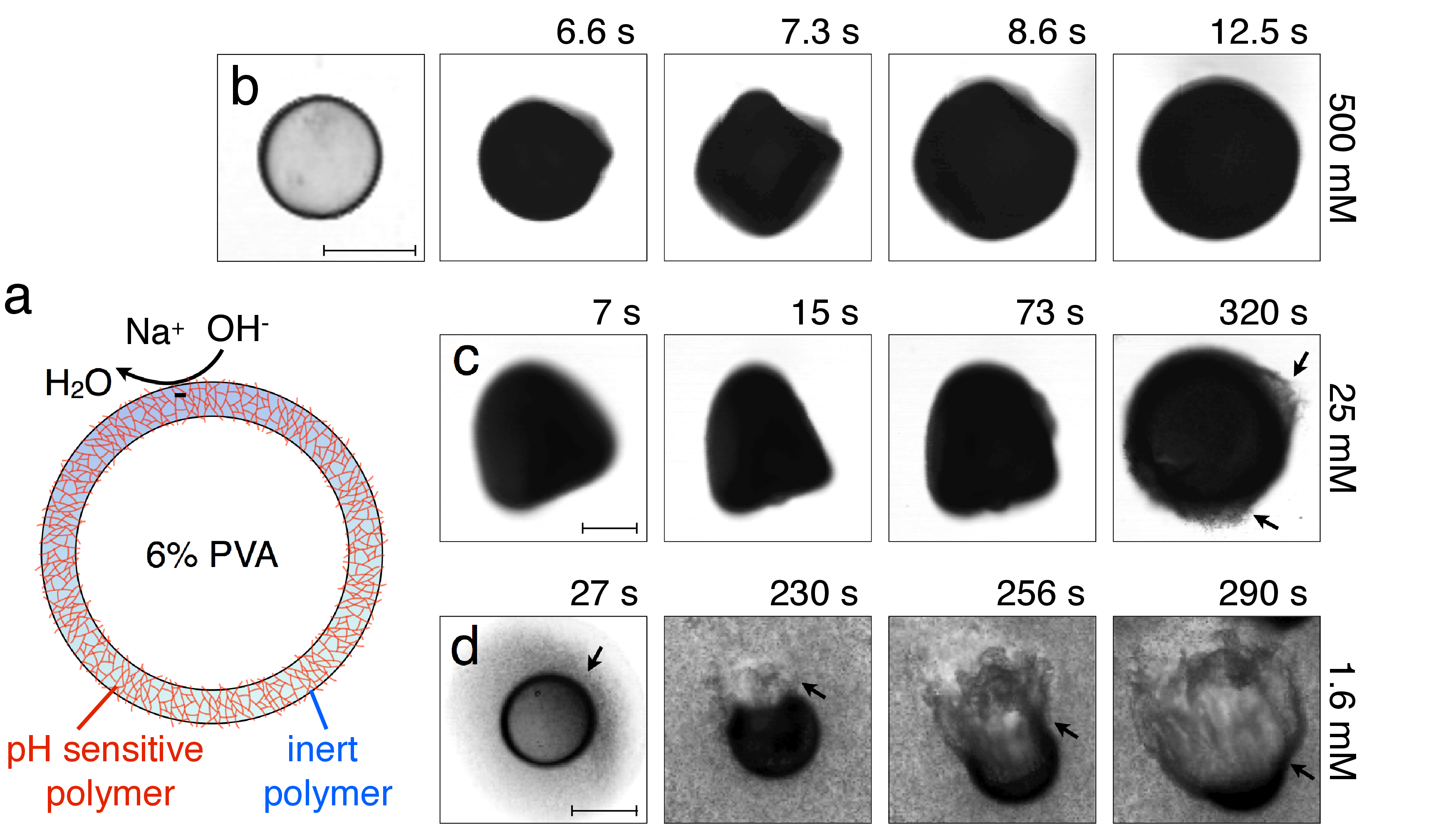}
\caption{(a) Schematic of a hybrid microcapsule, showing the thin, solid, spherical shell composed of a pH-unresponsive polymer and a pH-responsive polymer. The core is a 6 wt\% aqueous solution of polyvinyl alcohol (PVA). The pH-responsive polymer at the surface of the shell becomes charged when exposed to NaOH in the continuous phase. First frame of (b) shows an optical micrograph of a hybrid microcapsule after it is collected. Subsequent frames in (b), (c), and (d) show the expansion of the microcapsule when exposed to different NaOH concentrations. (c) At $c_{NaOH}=25~\mbox{mM}$, the shell begins to disintegrate, as indicated by the arrows in the last frame. (d) For lower $c_{NaOH}<25~\mbox{mM}$, the shell first disintegrates, as indicated by the arrow in the first frame, then ruptures through the formation of a hole, shown in the second frame, then expands into a wrinkled jellyfish-like structure, shown in the last two frames. Scale bars indicate 100 $\mu$m.} 
\end{center}
\end{figure*}

We fabricate monodisperse thin-shelled microcapsules using water-in-oil-in-water (W/O/W) double emulsion templates prepared by microfluidics.\cite{utada,supp} The inner and outer phases are 6 wt \% and 10 wt \% solutions of polyvinyl alcohol (PVA), respectively, while the middle oil phase is a mixture of a pH-responsive polymer, suspended in tetrahydrofuran (THF), and another photo-polymerizable pH-unresponsive monomer. We use UV light to polymerize the pH-unresponsive monomer {\it in situ}, immediately as the double emulsions are generated, forming a highly cross-linked network around the inner core. This network is a solid characterized by a Young's modulus $E\approx600~\mbox{MPa}$;\cite{lin} it is impermeable to hydrated Na$^{+}$ and OH$^{-}$ ions, but is permeable to water.\cite{shin1,dattabuckling} We then collect the microcapsules in water adjusted to have pH = 6, and let the THF evaporate; this forces the pH-responsive polymer to precipitate, completing the formation of a solid, hybrid shell,\cite{softmatter} schematized in Figure 1(a). We then remove the supernatant, and repeatedly wash the microcapsules with pure water, to remove any surfactant from the continuous phase.

When exposed to NaOH, the pH-responsive polymer chains at the microcapsule exteriors become highly charged and repel each other; this repulsion is screened by any residual Na$^+$ or OH$^-$ ions. To probe the microcapsule deformations under these conditions, we immerse them in an aqueous solution with $c_{NaOH}=500~\mbox{mM}$, and monitor them using optical microscopy. Strikingly, the hybrid microcapsules quickly become opaque, reflecting the development of heterogeneities in the shell, and abruptly expand in irregular shapes; representative optical micrographs of this process are shown in Figure 1(b) and Movie S1. The microcapsules ultimately expand into spheres, as shown by the last frame of Figure 1(b); this entire process occurs over a timescale of $\sim10~\mbox{s}$.

We quantify this behavior by measuring the maximal expansion, $\gamma\equiv(R_{f}-R_{0})/R_{0}$, where $R_{0}$ and $R_{f}$ are the average initial and final microcapsule radii, respectively. We also measure the average time taken for the microcapsules to expand into their final spherical shapes, $\tau$. To elucidate the expansion behavior, we explore even lower values of $c_{NaOH}$. Intriguingly, $\gamma$ increases strongly with decreasing $c_{NaOH}$, as shown in the upper panel of Figure 2, reaching a value exceeding 1 at $c_{NaOH}=25~\mbox{mM}$; the expansion time $\tau$ concomitantly increases, but only weakly, as shown by the grey squares in the lower panel of Figure 2. Unexpectedly, at $c_{NaOH}=25~\mbox{mM}$, the microcapsules also begin to disintegrate, forming fragments of size $\sim1~\mu\mbox{m}$, as indicated by the arrows in Figure 1(c). At even smaller $c_{NaOH}$, the microcapsules first disintegrate, as shown by the first frame of Figure 1(d); this ultimately results in the formation of a hole in each shell after a time delay $\tau_{1}\sim10-100~\mbox{s}$. The microcapsules then begin to intermittently expand outward, starting at the hole perimeters, as exemplified by the last three frames of Figure 1(d) and Movie S2; the region of the shell that expands is indicated by the arrow in each frame. This process occurs over a time period $\tau_{2}\sim10-100~\mbox{s}$; the expansion then stops, leaving wrinkled jellyfish-like structures, approximately $300~\mu$m in size, as shown in Movie S2. 

The isotropic expansion at high $c_{NaOH}\geq25~\mbox{mM}$ is due to the charging of the microcapsule shells; when exposed to NaOH, the pH-responsive polymers at the microcapsule exteriors become highly charged, leading to a surface charge density $\sigma$. We thus expect that the repulsion between the charges on each spherically-symmetric shell leads to a force, directed radially outward, on the shell. We estimate\cite{virus1,supp} the magnitude of the resultant electrostatic pressure as $p_{e}\approx\sigma^{2}\kappa^{-1}/\epsilon\epsilon_{0}R$, where $\epsilon\approx80$ is the dielectric constant of water, $\epsilon_{0}$ is the permittivity of free space, $R$ is the time-dependent radius of the spherical microcapsule, $\kappa^{-1}$ is the length over which the repulsive interactions are screened. The quantity of NaOH required to fully charge the shell is small;\cite{supp} consequently, the excess Na$^{+}$ and OH$^{-}$ ions, which screen the surface charges on the shell, have a concentration $\approx c_{NaOH}$, and thus $\kappa^{-1}=A/\sqrt{c_{NaOH}}$, with $A\equiv0.304~\mbox{nm}\cdot\mbox{M}^{1/2}$.\cite{israel} As the microcapsule expands isotropically, a tensile stress builds up within its shell; this resists the expansion. We use shell theory to estimate this stress, $p_{m}=Eh(1/R_{0}-1/R)$, where $h\approx3~\mu\mbox{m}$ is the average shell thickness, measured using scanning electron microscopy.\cite{landau} We thus expect that the microcapsules expand until the electrostatic pressure and tensile stress balance each other; this yields a maximal expansion strain $\gamma=\sigma^{2}A/E\epsilon\epsilon_{0}h\sqrt{c_{NaOH}}$. We test this prediction using our measurements of $\gamma$. We find $\gamma\sim c_{NaOH}^{-1/2}$, as shown by the line in the top panel of Figure 2, consistent with the theoretical prediction; fitting the experimental data yields an estimate of $\sigma\approx300~e/\mbox{nm}^{2}$, where $e$ is the charge of an electron. These results suggest that the microcapsule expansion can be understood by coupling electrostatics with shell theory. 
 \begin{figure}
\begin{center}
\includegraphics[width=3.7in]{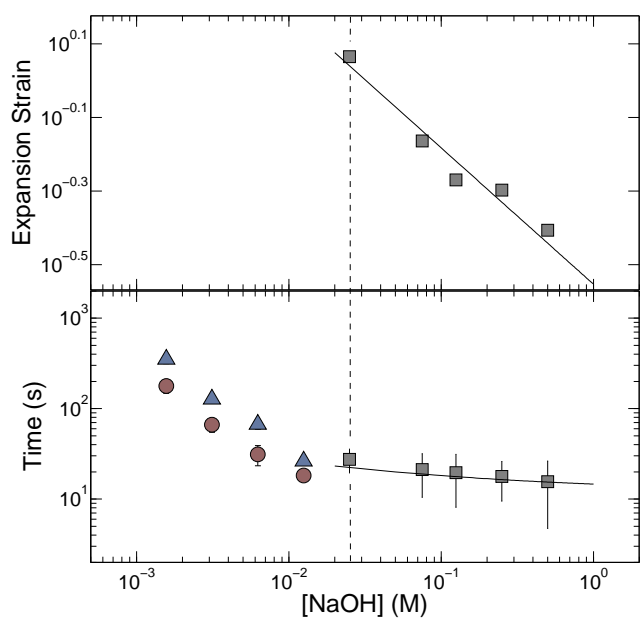}
\caption{(Top Panel) Maximal expansion strain of spherical microcapsules exposed to different NaOH concentrations; line shows theoretical prediction described in the text, with $\sigma\approx300~e/\mbox{nm}^{2}$, where $e$ is the elementary charge. (Bottom Panel) Grey squares show time taken for spherical microcapsules to fully expand; line shows theoretical fit described in the text. Blue triangles and red circles the time period over which the shell disintegrates before a hole forms in it, $\tau_{1}$, and the time period over which the shell subsequently expands into a jellyfish-like structure, $\tau_{2}$, respectively, for $c_{NaOH}<25~\mbox{mM}$.} 
\end{center}
\end{figure}

Within this picture, the Na$^{+}$ and OH$^{-}$ ions screen the repulsion between the charges on the microcapsule shells. As an additional test of this idea, we expose the microcapsules to a 25 mM NaOH solution, also containing 500 mM NaCl; we expect the Na$^{+}$ and Cl$^{-}$ ions to similarly screen the repulsion between the charges on the microcapsule shells. Consistent with our expectation, the microcapsules do not disintegrate, unlike microcapsules exposed to only 25 mM NaOH. Instead, they expand slightly, reaching a maximal expansion strain $\gamma\approx0.4$, much smaller than that expected for $c_{NaOH}$. This value of the strain is instead comparable to that expected for $c_{NaOH}=525~\mbox{mM}$, as shown by the line in the top panel of Figure 2, thus confirming the validity of our picture.

The microcapsules expand to their maximal size over a time period $\tau$. We hypothesize that this behavior reflects the dynamics of the fluid flow into the microcapsule cores, through their porous shells, as they expand.\cite{dattabuckling} We estimate this inflow rate using Darcy's law, $p_{e}-p_{m}=\mu h\dot{R}/3k$, where $\mu\approx1~\mbox{mPa}\cdot\mbox{s}$ is the viscosity of water and $k$ is the shell permeability; this yields a characteristic expansion time $\tau=\mu R_{0}^{2}(1+\gamma)/3Ek$. We use our measurements of $\tau$, as well as the fit to the measurements of $\gamma$ shown in the top panel of Figure 2, to directly test this prediction. We find excellent agreement between the data and the theoretical prediction, as shown by the solid line in the bottom panel of Figure 2; fitting the experimental data yields a shell permeability $k\approx10^{-21}~\mbox{m}^{2}$. The agreement between the data and the theoretical prediction thus suggests that the dynamics of the microcapsule expansion can be understood using Darcy's law.

As $c_{NaOH}$ decreases, the screening length $\kappa^{-1}$ increases; consequently, the stress in the microcapsule shell increases, ultimately reaching $p_{m}=p_{e}\approx10~\mbox{MPa}$ at $c_{NaOH}=25~\mbox{mM}$. For even smaller $c_{NaOH}$, the microcapsules begin to disintegrate into $\sim1~\mu\mbox{m}$ fragments, and eventually rupture. We hypothesize that, under these conditions, the stress that builds up in the shell exceeds the stress required to fracture it. To test this hypothesis, we estimate the fracture stress\cite{griffiths} using the Griffith criterion, $\sqrt{2EG_{c}/\pi a}\approx6-20~\mbox{MPa}$, where $G_{c}\sim0.1-1~\mbox{J}/\mbox{m}^{2}$ is the surface energy per unit area of the shell material and $a\approx1~\mu\mbox{m}$ is the characteristic size of a shell fragment, measured using optical microscopy. Interestingly, this value is in good agreement with our estimate of the maximal stress that develops in the shell, $p_{m}\approx10~\mbox{MPa}$, suggesting that the observed disintegration at $c_{NaOH}\leq25~\mbox{mM}$ reflects the fracturing of the microcapsule shell. 

Finally, we monitor the dynamics of the disintegration, rupture, and subsequent expansion of the microcapsule shell that occurs at $c_{NaOH}<25~\mbox{mM}$. Intriguingly, both the time period over which the shell fractures before a hole forms in it, $\tau_{1}$, and the time period over which the shell subsequently expands into a jellyfish-like structure, $\tau_{2}$, both increase with decreasing $c_{NaOH}$, as shown by the blue triangles and red circles in Figure 2, respectively. A complete understanding of these dynamics remains a puzzle requiring further inquiry. 

Our work highlights the fascinating range of structures exhibited by pH-responsive microcapsules, driven by the interplay between electrostatic and mechanical forces. Intriguingly, many of the morphologies we observe are reminiscent of structures, also induced by electrostatic effects, that often occur in other soft matter systems, such as viruses\cite{virus1,virus2,hole2} and red blood cells\cite{hole1}.

It is a pleasure to acknowledge Esther Amstad, Michael Brenner, Alberto Fernandez-Nieves, Rodrigo Guerra, and Zhigang Suo for stimulating discussions. This work was supported by the NSF (DMR-1006546) and the Harvard MRSEC (DMR-0820484). SSD acknowledges funding from ConocoPhillips. AA acknowledges Evonik Industries for generously donating the Eudragit S-100 polymer used.




\footnotesize{
\bibliography{rsc} 
\bibliographystyle{rsc} }

\end{document}